\documentclass[pre,showpacs,superscriptaddress,fleqn,twocolumn]{revtex4}
\usepackage{graphicx}
\usepackage{latexsym}
\usepackage{amsfonts}
\usepackage{amssymb}
\usepackage{amsmath}
\usepackage{bbm}
\usepackage{xcolor}

\newcommand{\be}{\begin{equation}}
\newcommand{\ee}{\end{equation}}
\newcommand{\bel}[1]{\begin{equation}\label{#1}}
\newcommand{\bea}{\begin{eqnarray}}
\newcommand{\eea}{\end{eqnarray}}
\newcommand{\bef}{\begin{figure}}
\newcommand{\enf}{\end{figure}}
\newcommand{\ba}{\begin{array}}
\newcommand{\ball}{\begin{array}{ll}}
\newcommand{\bacl}{\begin{array}{cl}}
\newcommand{\bacll}{\begin{array}{cll}}
\newcommand{\bal}{\begin{array}{l}}
\newcommand{\bac}{\begin{array}{c}}
\newcommand{\ea}{\end{array}}
\newcommand{\N}{{\mathbb{N}}}
\newcommand{\R}{{\mathbb{R}}}
\newcommand{\E}{{\mathbb{E}}}
\renewcommand{\P}{{\mathbb{P}}}

\begin{document}

\title{Instability of condensation in the zero-range process with random interaction}

\author{Stefan Grosskinsky}
\affiliation{Mathematics and Complexity Science, University of Warwick, Coventry CV4 7AL, UK}
\author{Paul Chleboun}
\affiliation{Mathematics and Complexity Science, University of Warwick, Coventry CV4 7AL, UK}
\author{Gunter M.~Sch\"utz\footnote{This work was partially carried out while GMS was 
Weston Visiting Professor at the Weizmann Institute of Science, Israel}}
\affiliation{Institut f\"ur Festk\"orperforschung, Forschungszentrum J\"ulich, D-52425
J\"ulich, Germany}

\date{\today }

\begin{abstract} The zero-range process is a stochastic interacting particle system that is known to exhibit a condensation transition. We present a detailed 
analysis of this transition in the presence of quenched disorder in the particle 
interactions.  Using rigorous probabilistic arguments we show that disorder changes 
the critical exponent in the interaction strength below which a condensation transition may occur. 
The local critical densities may exhibit large fluctuations and their distribution shows an interesting crossover from exponential to algebraic behaviour.

\end{abstract}
\pacs{05.40.-a, 02.50.Ey, 64.60.Cn}

\maketitle

\definecolor{dyellow}{rgb}{0.5,0.5,0}

The zero-range process is a stochastic lattice gas where the particles hop randomly
with an on-site interaction that makes the jump rate dependent only on the local
particle number.  It was introduced in \cite{spitzer70} as a mathematical model for interacting
diffusing particles, and since then has been applied in a large variety of contexts, 
often under different names, (see e.g. \cite{evansetal05} and references therein). 
The model is simple enough for the steady state to factorize, on the other hand it 
exhibits an interesting condensation transition under certain conditions. Viz. when 
the particle density exceeds a critical value $\rho_c$ the system phase separates 
into a homogeneous background with density $\rho_c$ and all the excess mass 
concentrates on a single lattice site. This has been observed and studied in some detail in 
experiments on shaken granular media 
\cite{meeretal02,meeretal07}, and there is a well-established analogy
to Bose-Einstein condensation \cite{evansetal05,krugetal96,evansetal06}.
It is also relevant as a generic mechanism for phase separation in single-file diffusion 
\cite{kafrietal02} and condensation phenomena in many complex systems such as network 
rewiring \cite{angeletal06} or traffic flow \cite{Kaup05}, 
for a review see \cite{evansetal05}.

The transition can be caused by site-dependent jump rates $g_x$
\cite{krugetal96} due to the slowest site acting as a trap.
It also appears in a more subtle fashion in homogeneous 
systems where condensation may result from
the growth of large clusters on the expense of small clusters,
if the jump rates $g(n)$ as a function of the 
number $n$ of particles on the starting site have a decreasing tail.
Such a model with a generic power law decay
\bea\label{rates}
g(n)=1+b/n^\sigma 
\eea
with positive interaction parameters $b,\sigma$ has been
 introduced in \cite{evans00}. Condensation occurs if 
$0<\sigma <1$ and $b>0$ or if $\sigma =1$ and $b>2$. 
The condensation transition is understood also rigorously in the context of the equivalence of ensembles \cite{stefan,evansetal05b}, and more 
recently many variants of (\ref{rates}) have been studied 
\cite{evansetal05,schwarzkopfetal08,lucketal07,angeletal07,stefan2,evansetal06}.
\bef
\begin{center}
\includegraphics[width=0.4\textwidth]{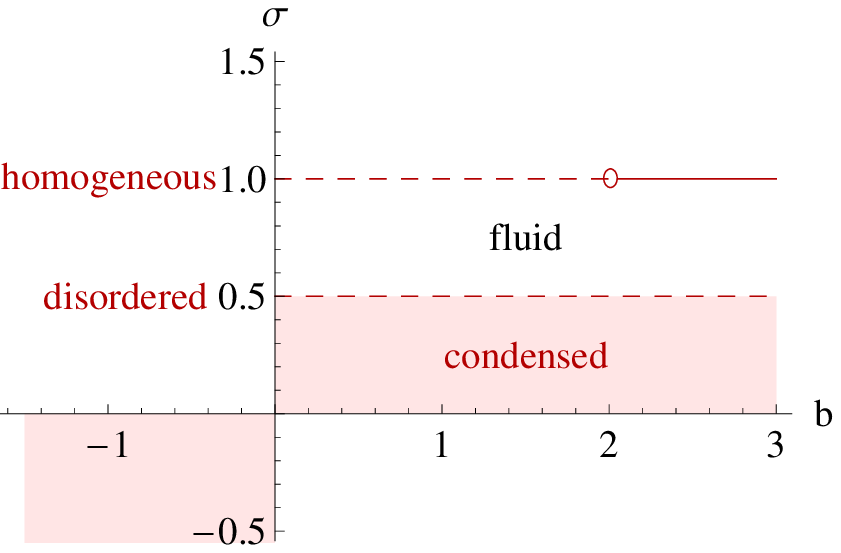}
\end{center}
\caption{\label{fig1}
(Color online) Change of the phase diagram under random perturbations of the jump rates (\ref{rates}).
Disorder changes the critical interaction exponent from 1 to 1/2 and leads to a critical density
that depends on the system size. Inside the red-shaded region condensation occurs above a non-zero critical density for $b>0$, and for $b<0$ (see (\ref{rates2}))
and negative $\sigma$ the critical density is zero.}
\enf

All previous studies of zero-range processes assume that the interaction between particles is strictly equal on all sites. 
In this paper we study the effect of disorder on this interaction 
and show that even a small random perturbation of the $n$-dependence of the
jump rates $g(n)$ leads to a drastic change in the critical behaviour. Namely, for positive $b$ condensation only occurs for $0<\sigma <1/2$ (see Figure \ref{fig1}), excluding in particular the frequently studied case $\sigma =1$. Moreover, the critical densities are site-dependent random variables with non-trivial distributions and depending on the parameter values they may exhibit large fluctuations.
Due to the wide applicability of the zero-range process, the change of 
the critical interaction exponent $\sigma$ is particularly relevant for 
applications as is explained in the conclusion.

We consider a lattice $\Lambda_L$, which we take to be periodic and of finite 
size $|\Lambda_L |=L$. A configuration is denoted by $(\eta_x )_{x\in\Lambda}$ 
where $\eta_x \in\{ 0,1,\ldots\}$ is the occupation number at site $x$. The 
dynamics is defined in continuous time, such that with rate $g_x (\eta_x )$ site 
$x\in\Lambda_L$ loses a particle, which moves to a randomly chosen target 
site $y$ according to some probability distribution $p(y{-}x)$. For 
example in one dimension with nearest neighbour hopping, the particle 
moves to the right (left) with probability $p$ ($1{-}p$). 

A generic perturbation of the jump rates (\ref{rates}) can be additive or multiplicative, but since the condensation behaviour is determined only by the tail of the jump rates for large $n$, both choices are essentially equivalent.  They can be written in a convenient general way,
\bea\label{rates2}
g_x (n)=e^{E_x (n)} \quad\mbox{for }n\geq 1\ ,\quad g(0)=0\ ,
\eea
where the exponents are given by
\bea\label{exk}
E_x (n)=e_x (n) +b/n^\sigma\ ,\quad b\in\R ,\ \sigma >0\ ,
\eea
with $e_x (n)$ being iid random variables with respect to $x$ and $n$. Without $x$-dependence this would amount merely to a change of the function $g$ which might be interesting, but is a degenerate problem in terms of generic perturbations. On the other hand, the effect of spatially inhomogeneous jump rates favouring condensation on slow sites has already been studied \cite{krugetal96}. Therefore we concentrate on disorder with spatially uniform mean $\E\big( e_x (n)\big) =0$ and variance $\delta^2 >0$, in order to focus on the basic novelty which is the supressing effect on condensation for a generic perturbation of (\ref{rates}). For the same reason we have chosen the jump probability $p(y{-}x)$ to be homogeneous, since spatial dependence there leads to the same effect as spatially inhomogeneous jump rates \cite{comment}.\\
Note that for $e_x (n)\equiv 0$ (i.e. $\delta =0$) the asymptotic behaviour of $g_x (n)$ is given by (\ref{rates}).  
All data shown in this paper are for uniform $e_x (n)\sim U(-\alpha ,\alpha )$, characterized by $\delta^2 =\alpha^2 /3$. But our analytical results are of course independent of the distribution of the perturbation as well as the exact form of the jump rates (\ref{rates2}). For negative $\sigma$ the rates (\ref{rates2}) are increasing in $n$ for positive $b$ and hence there is no condensation. For negative $b$ the rates tend to zero, which means that there is condensation at critical density  $\rho_c=0$. This is an essentially trivial feature of the model which is robust against perturbation by disorder. We therefore focus on positive interaction exponent $\sigma$.

It is well known (see e.g. \cite{andjel82,evansetal05}) that the process has a 
grand-canonical factorized steady state $\nu_\mu^L$ with single-site marginal
\bea\label{marginal}
\nu_{x,\mu} (n)&=&\frac{e^{n\,\mu}}{z_x (\mu )}\prod_{k=1}^n g_x (k)^{-1} =\nonumber\\
&=&\frac{1}{z_x (\mu )} \exp\Big( n\,\mu -\sum_{k=1}^n E_x (k)\Big)\ ,
\eea
where the chemical potential $\mu\in\R$ fixes the particle density. This holds 
independently of the distribution of target sites $p(y-x)$ and for each realization 
of the $e_x (k)$, i.e. $\nu_\mu^L$ is a \textit{quenched} distribution. The single-site normalization 
is given by the partition function
\bea\label{zxmu}
z_x (\mu )=\sum_{n=0}^\infty \exp\Big( n\,\mu -\sum_{k=1}^n E_x (k)\Big)
\eea
which is strictly increasing and convex in $\mu$ \cite{stefan}. The local density can be 
calculated as usual as a derivative of the free energy
\bea\label{rhox}
\rho^x (\mu )=\langle\eta_x \rangle =\frac{\partial \log z_x (\mu )}{\partial\mu}
\eea
and is a strictly increasing function of $\mu$. By $\langle ..\rangle$ we denote the 
(quenched) expected value with respect to $\nu_\mu^L$ for fixed disorder, i.e. 
fixed realization of the $e_x (k)$.
To study the condensation transition we have to identify the maximal or 
critical chemical potential $\mu_c \in\R$, such that $z_x (\mu )<\infty$ for 
all $\mu <\mu_c$. 

\bef[t]
\begin{center}
\includegraphics[width=0.4\textwidth]{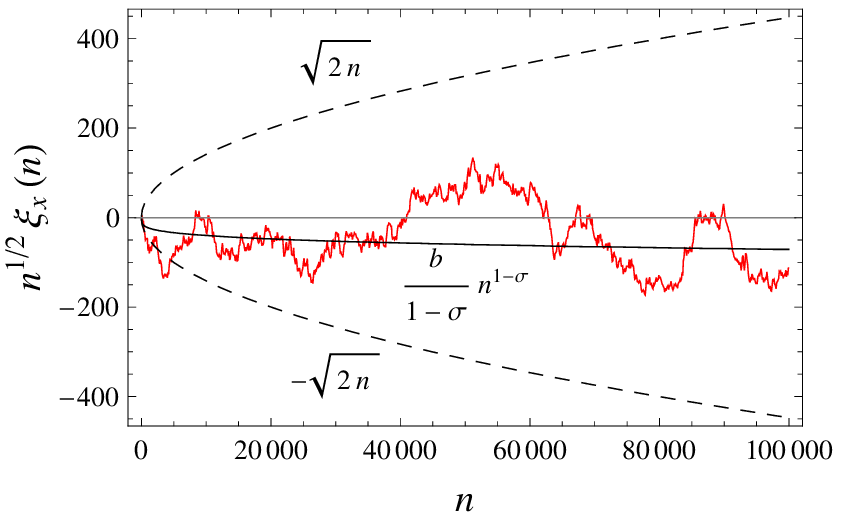}
\end{center}
\caption{\label{fig1b}
(Color online) Typical realizations of $\sqrt{n}\,\xi_x (n)$ (drawn with $\delta^2 {=}1$) leave the area enclosed 
by the dashed parabola only finitely many times, but cross the full line
$-\frac{b}{1-\sigma}\, n^{1-\sigma}$ for $\sigma >1/2$ infinitely often (here $\sigma =0.75$, $b=1$).
}
\enf
The single-site marginal (\ref{marginal}) is a function of the disorder and for 
$b=0$ it has the distribution of a geometric random walk with deterministic 
drift $\mu$. 
The critical chemical potential in this case is simply $\mu_c =0$ and as 
$n\to\infty$ the distribution $\nu_{x,\mu} (n)$ converges to a log-normal.
With non-zero $b$ the drift term becomes $n$-dependent and a 
more detailed analysis is required. With (\ref{exk}) we have to leading 
order as $n\to\infty$
\bea\label{exkn}
\sum_{k=1}^n E_x (k)\simeq\delta\sqrt{n}\,\xi_x (n)+
\left\{\bacl \frac{b}{1-\sigma} n^{1-\sigma} \ &,\ \sigma\neq 1\\ b\ln n\ &,\ \sigma =1\ea\right.\ ,
\eea
where by the central limit theorem
\bea\label{clt}
\xi_x (n):=\frac{1}{\delta\sqrt{n}} \sum_{k=1}^n e_x (k)\stackrel{n\to\infty}{\longrightarrow} N(0,1)
\eea
converges to a standard Gaussian. Moreover, the process 
$\big(\sqrt{n}\,\xi_x (n):n\in\N\big)$ is a random walk with increments of 
mean zero and variance $1$. Since the fluctuations of such a process are 
of order $\sqrt{n}$ we have for all $C\in\R$
\bea\label{argu}
\P \big(\xi_x (n)\lessgtr C\mbox{ for infinitely many }n\big) =1\ ,
\eea
and for all $\gamma >0,\ C>0$
\bea\label{argu2}
\P \big(|\xi_x (n)|>Cn^\gamma\mbox{ for infinitely many }n\big) =0\ .
\eea
This is a direct consequence of the law of the iterated logarithm
(see e.g. \cite{kallenberg}, Corollary 14.8)
and is illustrated in Figure \ref{fig1b}. 
Together with (\ref{zxmu}) and (\ref{exkn}) this implies that $z_x (\mu )<\infty$ for all $\mu <0$ with probability one. So for almost all (in a 
probabilistic sense) realizations of the $e_x (k)$ the critical
chemical potential is $\mu_c =0$ and
\bea\label{zxmuc}
z_x (\mu_c )=\sum_{n=0}^\infty \exp\bigg( -\sum_{k=1}^n \Big( e_x (k)+\frac{b}{k^\sigma}\Big)\bigg)\ .
\eea
For certain values of $\sigma$ and $b$, $z_x (\mu_c )<\infty$ is possible 
and $\nu_{x,\mu_c }$ can be normalized, which is a necessary condition 
for a condensation transition \cite{evans00}. We find

\textbf{$b\leq 0$:} In this case (\ref{exkn}) and (\ref{argu}) imply that there are infinitely many terms in 
(\ref{zxmuc}) which are bounded below by $1$. Since all terms of the sum are non-negative 
it diverges and $z_x (\mu_c )=\infty$ with probability one.

\textbf{$b>0$:} In this case the asymptotic behaviour of the terms in the sum (\ref{zxmuc}) 
depends on the value of $\sigma >0$, since the sign of the exponent can change.
\begin{itemize}
\item For $\sigma >1/2$, $n^{1-\sigma} \ll\sqrt{n}$ and (\ref{exkn}) is 
dominated by $\delta\sqrt{n}\,\xi_x (n)$. Applying (\ref{argu}) with 
$C=0$ we get $z_x (\mu_c )=\infty$ with probability one.
\item For $\sigma =1/2$ both terms in (\ref{exkn}) are of the same order since $n^{1-\sigma} \sim\sqrt{n}$ and
	\bea
	{-}\sum_{k=1}^n \Big( e_x (k){+}\frac{b}{k^\sigma}\Big)\simeq {-}\delta\sqrt{n}\big(\xi_x (n){+}\tfrac{2b}{\delta}\big)\ .
	\eea
Again, (\ref{argu}) this time with $C=2b/\delta$ implies $z_x (\mu_c )=\infty$ with probability one.
\item For $0<\sigma <1/2$ we have $n^{1-\sigma} \gg\sqrt{n}$ and 
(\ref{exkn}) is dominated by $\frac{b}{1-\sigma} n^{1-\sigma}$. We apply 
(\ref{argu2}) for $\gamma =1-\sigma -1/2>0$ to see that the 
random quantity $\xi_x (n)$ can change the sign of the exponent in 
(\ref{zxmuc}) only for finitely many terms in the sum. Therefore $z_x (\mu_c )<\infty$ 
with probability one since $\frac{b}{1-\sigma} n^{1-\sigma}$ has a fixed negative 
sign in (\ref{zxmuc}).
\end{itemize}

Whenever $z_x (\mu_c )=\infty$ the local critical density
\bea
\rho_c^x :=\rho^x (\mu_c )=\infty
\eea
also diverges and there is no condensation transition (see e.g. \cite{kipnislandim}, Lemma I.3.3). 
But for $b>0$ and $0<\sigma <1/2$ we have $z_x (\mu_c )<\infty$ and 
by the same argument as above it follows that
\bea\label{rhocx}
\rho_c^x =\frac{1}{z_x (\mu_c )} \sum_{n=0}^\infty n\, e^{ -\sum_{k=1}^n ( e_x (k)+\frac{b}{k^\sigma})} <\infty
\eea
with probability one, since the factor $n$ in the sum only gives a logarithmic 
correction in the exponent. Therefore there is condensation since
the grand-canonical product measures only exist
up to a total density of
\bea\label{rhocrit}
\rho_c (L):=\frac1L\sum_{x\in\Lambda_L} \rho_c^x \ ,
\eea
which depends on the $e_x (n)$ and the size of the lattice $L$.

\bef
\begin{center}
\includegraphics[width=0.4\textwidth]{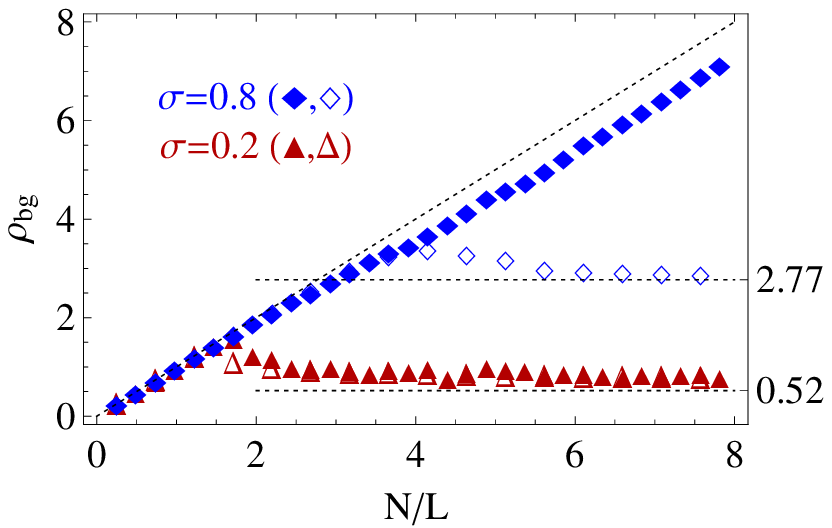}
\end{center}
\caption{\label{fig3}
(Color online) Background density $\rho_{bg}$ (\ref{rhobg}) as a function of $N/L$ for $b=1.2$. Data for fixed disorder with $\delta^2 =1/12$ and $L=1024$ show condensation for $\sigma =0.2$ and no condensation for $\sigma =0.8$. Data for $\delta^2 =0$ are shown by unfilled symbols, dotted lines indicate the critical densities. Data points are averages over 100 MC samples with errors of the size of the symbols.
}
\enf

If the actual number of particles $N$ is larger than $L\,\rho_c (L)$, 
all sites except the `slowest' one contain on average 
$\langle\eta_x \rangle_N =\rho_c^x$ particles. They form the so-called 
critical \textit{background}, since their distribution has non-exponential tails. By $\langle ..\rangle_N$ we denote the (canonical) expectation conditioned on the total particel number $N$.
The slowest site, say $y$, is defined by $\rho_c^y >\rho_c^x$ for all $x\neq y$.
By the conservation law it is required that
\bea\label{ordern}
\langle\eta_y \rangle_N =N-\sum_{x\neq y} \rho_c^x =O(N)\ ,
\eea
i.e. it contains of order $N$ particles and forms the \textit{condensate}. This 
interpretation is in accordance with previous results and has been proved 
rigorously in \cite{ferrarietal07} in the limit $N\to\infty$ for the unperturbed 
model. For the perturbed model we support this conclusion by MC simulations, some of
which are shown in Figure \ref{fig3}. For fixed $L$ we plot the stationary background density
\bea\label{rhobg}
\rho_{bg} :=\frac1L \big( N-\langle\eta_y \rangle_N \big)
\eea
as a function of the total density $\rho =N/L$. For $\sigma =0.2$, $\rho_{bg}$ converges to a critical density $\rho_c (L)$ which is slightly higher than for the unperturbed model. The overshoot of $\rho_{bg}$ for densities close to $\rho_c (L)$ is due to sampling from the canonical rather than the grand canonical ensemble, which has been observed already in \cite{angeletal04}. For $\sigma =0.8$, $\rho_{bg}$ increases approximately linearly with $\rho$, which is clearly different from the unperturbed model condensing with critical density $2.77$. 

For the perturbed system the critical density $\rho_c (L)$ is a random variable, which according to (\ref{rhocrit}) converges in the thermodynamic limit to the expected value $\E (\rho_c^x )$ with respect to the disorder.
This is in general hard to calculate \cite{yor}, even for simple choices of the $e_x (n)$. Detailed numerical estimates of the distribution of $\rho_c^x$ (some of which are shown in Fig. \ref{fig2b}) indicate that depending on the system parameters the cumulative tail is either algebraic or consists of an algebraic part with an exponential cut-off at large values. This can be explained heuristically by the interplay of the two terms in (\ref{exkn}) that determine the main contributions to the partition function \cite{inprogress}. The exponents of the purely algebraic tails is smaller than but often close to $-1$ and the length scale of the exponential tails can be extremely large (cf. Fig. \ref{fig2b}). Therefore $\rho_c^x$ has a finite mean which determines the thermodynamic limit of (\ref{rhocrit})
\bea\label{thlim}
\rho_c :=\lim_{L\to\infty} \rho_c (L)=\E (\rho_c^x )<\infty\ ,
\eea
but for finite $L$ the $\rho_c^x$ can exhibit large fluctuations, resulting in high values for $\rho_c$ and slow convergence of (\ref{thlim}).

\bef[t]
\begin{center}
\includegraphics[width=0.4\textwidth]{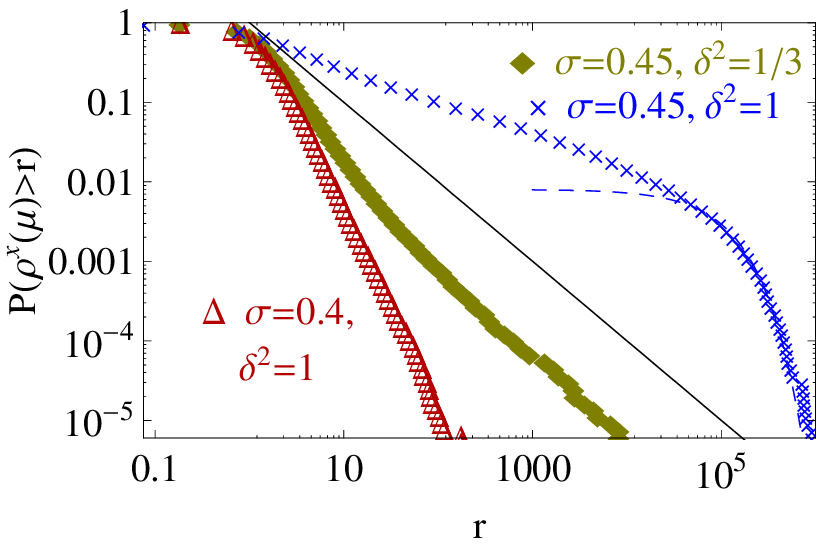}
\end{center}
\caption{\label{fig2b}
(Color online) Cumulative tail of the distribution of $\rho_c^x$ from $5\cdot 10^5$ independent numerical calculations of (\ref{rhocx}) with $b=1$, where infinite sums have been cut off at $n=10^6$. The full line corresponds to a cumulative exponent $-1$. The data exhibit power law decays and for some parameters also large exponential cut-offs, e.g. $\sim e^{-r/10^5}$ for $\sigma =0.45$, $\delta^2 =1$ fitted by a dashed line. The expectations are finite but may be very large, $\E (\rho_c^x )=1.60\, (\textcolor{red}{\Delta}),\ 2.55\, (\textcolor{dyellow}{\blacklozenge}),\ 1106\, (\textcolor{blue}{\times})$.
}
\enf

The interaction encoded in the
jump rates of a zero-range model represents an effective
interaction for which space-dependence and randomness
due to microscopic impurities or heterogeneities in the
case of complex systems has to be taken into account. 
We have shown that generic interaction
disorder reduces the critical interaction exponent from $\sigma=1$ to $\sigma=1/2$.
In particular, this implies that for the most-studied case $\sigma =1$ there is no condensation transition in the
presence of interaction disorder.
This case is relevant for the mapping of
the zero-range process to exclusion models
where it becomes an effective
model for domain wall dynamics and therefore a powerful
criterion for phase separation in more general particle
systems \cite{kafrietal02}. In this mapping
the spatial interaction disorder maps into hopping rates
which depend in a random fashion on the interparticle
distance. Then our results imply that in such heterogeneous
finite systems the change of the critical interaction exponent
has to be taken into account and finite size
effects may play a major role due to large fluctuations of
local critical densities.


\begin{thebibliography}{10}

\bibitem{spitzer70}
F.~Spitzer, 
Adv.\ Math. \textbf{5}, 246--290 (1970).

\bibitem{evansetal05}
M.R.~Evans, T.~Hanney, 
J.\ Phys.\ A: Math.\ Gen. \textbf{38}, R195--R239 (2005).

\bibitem{meeretal02}
J. Eggers, Phys. Rev. Lett. \textbf{83}(25), 5322-5325 (1999).
D.~van der Meer, J.P.~van der Weele, D.~Lohse,
Phys. Rev. Lett. \textbf{88}, 174302 (2002). F.~Coppex, M.~Droz, A.~Lipowski, Phys.~Rev.~E \textbf{66}, 011305 (2002).

\bibitem{meeretal07}
J.~T\"or\"ok,
Physica A \textbf{355}, 374--382 (2005). D.~van der Meer, K.~van der Weele, P.~Reimann, D.~Lohse, J. Stat. Mech.: Theor. Exp. P07021 (2007).


\bibitem{krugetal96}
M.R.~Evans,
Europhys. Lett. \textbf{36}, 13-18 (1996).
J.~Krug, P.A.~Ferrari, 
J.~Phys.~A: Math.~Gen.~\textbf{29}, L465--L471 (1996). 
I.~Benjamini, P.A.~Ferrari, C.~Landim, 
Stoch.~Proc.~Appl.~\textbf{61}, 181--204 (1996).

\bibitem{evansetal06}
M.R.~Evans, T.~Hanney, S.N.~Majumdar, 
Phys. Rev. Lett. \textbf{97}, 010602 (2006).

\bibitem{kafrietal02}
Y.\ Kafri, E.\ Levine, D.\ Mukamel, G.M.\ Sch\"utz, and J.\ T\"or\"ok.
\newblock Phys.\ Rev.\ Lett. \textbf{89}(3), 035702 (2002).

\bibitem{angeletal06}
A.G. Angel, M.R. Evans, E. Levine, D. Mukamel, Phys. Rev. E \textbf{72}, 046132 (2005). A.G. Angel, T. Hanney, M.R. Evans, Phys. Rev. E \textbf{73}, 016105 (2006).

\bibitem{Kaup05} J.~Kaupuzs, R.~Mahnke, R.J.~Harris,
Phys. Rev. E \textbf{72}(5), 056125 (2005).

\bibitem{evans00}
M.R.~Evans, 
Braz.\ J.\ Phys. \textbf{30}(1), 42--57 (2000).

\bibitem{stefan}
S.~Grosskinsky, G.M.~Sch\"utz, H.~Spohn, 
J.\ Stat.\ Phys. \textbf{113}(3/4), 389--410 (2003).

\bibitem{evansetal05b}
M.R.~Evans, S.N.~Majumdar, R.K.P.~Zia, 
J.\ Stat.\ Phys. \textbf{123}, 357--390 (2006).

\bibitem{lucketal07}
J.M.~Luck, C.~Godreche,
J. Stat. Mech.: Theor. Exp. P08005 (2007).

\bibitem{angeletal07}
A.G.~Angel, M.R.~Evans, E.~Levine, D.~Mukamel, 
J.~Stat.~Mech.: Theor. Exp. P08017 (2007).

\bibitem{stefan2}
S.~Grosskinsky, G.M.~Sch\"utz, 
J.~Stat.~Phys.~\textbf{132}(1), 77--108 (2008).

\bibitem{schwarzkopfetal08}
Y.~Schwarzkopf, M.R.~Evans, D.~Mukamel, 
J. Phys. A: Math. Theor. \textbf{41}, 205001 (2008).

\bibitem{comment}
Inhomogeneous $p(x,y)$ leads to multiplication of the fugacity $\phi =e^\mu$ in the steady state by site-dependent terms which are the unique solutions of $\phi_y =\sum_x \phi_x p(x,y)$.

\bibitem{andjel82}
E.~Andjel, 
Ann.\ Probability \textbf{10}(3), 525--547 (1982).

\bibitem{kallenberg}
O.~Kallenberg, {\em Foundations of Modern Probability, 2nd edition} (Springer, New York, 2002).

\bibitem{ferrarietal07}
P.A.~Ferrari, C.~Landim, V.V.~Sisko, 
J.~Stat.~Phys.~\textbf{128}, 1153-1158 (2007).

\bibitem{kipnislandim}
C.~Kipnis, C.~Landim, {\em Scaling Limits of Interacting Particle Systems}
(Springer, Berlin, 1999).

\bibitem{angeletal04}
A.G.~Angel, M.R.~Evans, D.~Mukamel,
J.~Stat.~Mech.: Theor. Exp. P04001 (2004).

\bibitem{yor}
The authors are only aware of results on integrals of exponential Brownian motions with constant drift, e.g. M.~Yor,
J.~Appl.~Prob. \textbf{29}, 202-208 (1992).

\bibitem{inprogress}
S.~Grosskinsky, P.~Chleboun, G.M.~Sch\"utz (in prep.)

\end{thebibliography}
\end{document}